\documentclass[prb,preprint,superscriptaddress,groupedaddress]{revtex4}
\usepackage{graphicx}
\usepackage{dcolumn}
\usepackage{bm}
\usepackage{comment}
\usepackage{color}
\usepackage{amsmath}
\usepackage{accents}
\usepackage{subfigure}

\newcommand{\be}{\begin{equation}}
\newcommand{\ee}{\end{equation}}
\newcommand{\bea}{\begin{eqnarray}}
\newcommand{\eea}{\end{eqnarray}}
\newcommand{\acosh}{\mathrm{acosh}}

\newcommand{\sn}{\mathrm{sn}}
\newcommand{\cn}{\mathrm{cn}}
\newcommand{\dn}{\mathrm{dn}}

\begin{document}
\title{Thermally-Assisted Spin-Transfer Torque Dynamics in Energy Space}
\author{D.~Pinna}
\email{daniele.pinna@nyu.edu}
\author{A.~D.~Kent}
\affiliation{Department of Physics, New York University, New York, NY 10003, USA}
\author{D.~L.~Stein}
\affiliation{Department of Physics, New York University, New York, NY 10003, USA}
\affiliation{Courant Institute of Mathematical Sciences, New York University, New
             York, NY 10012, USA}

\begin{abstract}
  We consider the general Landau-Lifshitz-Gilbert theory underlying
  the magnetization dynamics of a macrospin magnet subject to spin-torque
  effects and thermal fluctuations. Thermally activated dynamical
  properties are analyzed by averaging the full magnetization equations over constant-energy
  orbits. After averaging, all the relevant dynamical scenarios are a
  function of the ratio between hard and easy axis anisotropies. We derive
  analytically the range of currents for which limit cycles exist and
  discuss the regimes in which the constant energy orbit averaging
  technique is applicable.
\end{abstract}

\pacs{Valid PACS appear here}
\maketitle


\section{Introduction}

More than a decade has passed since the switching effects of spin-polarized
currents traversing nanomagnets were first explored
experimentally~\cite{Slon,Berger,Katine2000,Ozy}. The quest to characterize
how a current may be used to transfer spin-angular momentum into a magnetic
system has led to sweeping advances in the field of spintronics. The study
and development of spin-valves and magnetic tunnel junctions (see, for
example,~\cite{Brataas}) with the aim of constructing denser and more
efficient memory technologies is one important result of
spin-transfer-torque related research. Theoretical treatments usually begin
by treating the magnetization within the ferromagnetic film as a single
macrospin~\cite{Brown}. Spin-torque effects are then taken into account
phenomenologically by modifying the macrospin's dynamical
Landau-Lifshitz-Gilbert~(LLG) equation~\cite{Slon}. 

At the nanoscale size, however, one must extend the deterministic treatment
described above to include random forces due to thermal noise.  The effects
of noise become even more pronounced at low currents. A comprehensive
theoretical treatment therefore requires taking into account the interplay
between noise and the deterministic dynamics. The analysis of such effects
is complicated by the non-conservative nature of the spin-torque interaction.

The energetics underlying the thermally activated behavior of uniaxial
macrospins have been studied in simple situations both analytically and
numerically~\cite{APL,PRB,Taniguchi,Taniguchi2}, the latter is possible because of
significant computational improvements in graphics processing unit (GPU) technology. The advantage of
using a uniaxial macrospin model is that its thermally activated switching
dynamics can be modeled using a $1D$~stochastic differential
equation~(SDE), which is solvable via standard statistical methods. In
practical applications, though, magnets possess a hard axis away from which
magnetization deviations have an energy cost. This is particularly true in
thin film nanomagnets where, due to the shape of the sample, out of plane
magnetizations are energetically suppressed. 

Upon adding a hard axis anisotropy to the macrospin model, one is forced to
deal with three coupled SDEs. A complete solution is then obtainable only
numerically. Nonetheless, a useful approximation can be employed whenever
the conservative precessional timescale is small compared to the diffusion
timescale, the latter determined by the interplay between noise and spin
transfer dynamics. This approximation allows for the construction of a
constant-energy orbit averaged (CEOA) LLG~equation, which characterizes the
macrospin dynamics by focusing on its slow diffusion in energy space.

An early attempt that used the above approach consisted in writing down a
Fokker-Planck equation describing the energy diffusion of the biaxial
monodomain~\cite{Apalkov, Serpico}. Unfortunately, an energy diffusion
equation of this type remains almost as intractable as the original set of
dynamical equations. Nonetheless, Apalkov and Visscher~\cite{Apalkov}
estimated a linear exponential scaling ($\log\tau\propto (1-I)$) between
current and mean switching time for thermal switching. However, research work done 
on uniaxial monodomains has found that, in contrast to the
scaling form above, the dependence of mean switching time on current ---
written generally as $\log\tau\propto (1-I)^{\beta}$ --- depends on whether
one is studying a uniaxial or a biaxial macrospin model. For the former,
the exponent~$\beta$ was found to be 2 rather than 1~\cite{Taniguchi,APL,MMM}.

Alternatively, an effective Langevin~equation for the energy dynamics of a
uniaxial macrospin can also be contructed by averaging the magnetization dynamics over constant
energy trajectories~\cite{MMM}. Newhall and Vanden-Eijnden~\cite{Katie} numerically constructed an analogous
Langevin~equation for the energy dynamics of a biaxial macrospin and found
numerical evidence for the appearance of stable limit cycles in the
macrospin dynamics for certain current regimes.  Their analysis supported
the Apalkov-Visscher linear scaling result ($\beta=1)$~\cite{Apalkov}, and demonstrated a current
dependence of~$\beta$ for thermal switching starting from a stable limit
cycle state. These differing results implied that the transition between
the uniaxial and biaxial macrospin models had to be either continuous or at
most display a finite discontinuity.  Taniguchi~\cite{TaniguchiE}, by also
approximating the Fokker-Planck equation over constant energy trajectories,
showed how the exponent~$\beta$ differs from the result of Apalkov and
Visscher in the thermally activated regime, as well as from the results of
Newhall and Vanden-Eijnden for applied currents greater than a certain
critical value. Above this value, $\beta$~ceases to be constant and instead
varies nonlinearly as a function of applied current.

In this paper we extend the energy Langevin approach~\cite{MMM} in order to
study the biaxial macrospin model. We obtain analytical expressions for
three critical current values across which transitions between different
dynamical regimes occur. Two of these are found to be identical to those
independently derived by Taniguchi via different
means~\cite{TaniguchiE}. The third critical current sets a limit on the
current regimes for which CEOA is a valid approximation. Our analysis
explains the transition between the biaxial and uniaxial macrospin regimes,
thus shedding light on the contradictory conclusions obtained in the
previous literature.  We further apply our analytical approach to thermally
activated switching and obtain an analytical expression for the non-linear
exponential scaling discussed by Taniguchi~\cite{TaniguchiE}.


\section{General Formalism}

A simple model of a ferromagnet uses a Stoner-Wohlfarth monodomain
with magnetization $\mathbf{M}$ whose properties are fixed by its
shape anisotropy. The energy landscape experienced by $\mathbf{M}$ is
generally described by three terms: an applied field $\mathbf{H}$, an
easy-axis anisotropy energy $U_K$ with axis along $\mathbf{\hat{n}}_K$
(Fig.~1) in the $\mathbf{e}_x - \mathbf{e}_z$ plane making an angle
$\theta$ with the $\mathbf{e}_z$ axis, and a hard-axis anisotropy
energy $U_p$ with normal direction $\mathbf{\hat{n}}_D$ perpendicular
to $\mathbf{\hat{n}}_K$. The magnetization $\mathbf{M}$ is assumed to
be constant in magnitude and for simplicity normalized into a unit
direction vector $\mathbf{m}=\mathbf{M}/|\mathbf{M}|$. A
spin-polarized current $J$ enters the magnetic body in the
$-\mathbf{e}_y$ direction, with spin polarization $\eta$, and spin
direction along the $\mathbf{e}_z$ axis. The current exits in the same
direction, but with its average spin direction aligned
along~$\mathbf{M}$. The self-induced magnetic field of the current is
ignored here because the dimension $a$ is considered to be smaller
than $100$~$\mathrm{nm}$, where spin-current effects are expected to
become dominant over those due to the current-induced magnetic
field. The dynamics are described by the standard LLG~equation:

\begin{eqnarray}
\mathbf{\dot{m}}=&-&\gamma'\mathbf{m}\times\mathbf{H}_{\mathrm{eff}}-\alpha\gamma'\mathbf{m}\times\left(\mathbf{m}\times\mathbf{H}_{\mathrm{eff}}\right)\nonumber\\
&-&\gamma' j\mathbf{m}\times\left(\mathbf{m}\times\mathbf{\hat{n}}_p\right)+\gamma'\alpha j\mathbf{m}\times\mathbf{\hat{n}}_p,
\end{eqnarray}
where $\gamma'=\gamma/(1+\alpha^2)$ is the Gilbert~ratio, $\gamma$ is
the usual gyromagnetic ratio and $j=(\hbar/2e)\eta J$ is the
spin-angular momentum deposited per unit time with $\eta =
(J_{\uparrow}-J_{\downarrow})/(J_{\uparrow}+J_{\downarrow})$ the
spin-polarization of incident current $J$. The last two terms describe
a vector torque generated by current polarized in the direction
$\mathbf{\hat{n}}_p$. These are obtained by assuming that the
macrospin absorbs angular momentum from the spin-polarized current
only in the direction perpendicular to $\mathbf{m}$.~\cite{Slon}
\newline

To write $\mathbf{H}_{\mathrm{eff}}$ explicitly, we must construct an
energy landscape for the magnetic body. There are three main components
that need to be considered: a uniaxial anisotropy energy $U_K$, a
hard-axis anisotropy $U_P$ and a pure external field interaction
$U_H$. These are written as follows:

\begin{eqnarray}
U_K&=&-K (\mathbf{\hat{n}}_K\cdot\mathbf{m})^2\\
U_P&=&K_P(\mathbf{\hat{n}}_D\cdot\mathbf{m})^2\\
U_H&=&-M_S V \mathbf{m}\cdot\mathbf{H}_{ext}
\end{eqnarray}
In these equations, $M_S$ is the saturation magnetization, $K_P$ is
the hard-axis anisotropy, $K=(1/2) M_S V H_K$, $H_K$ is the
Stoner-Wohlfarth switching field (in units of Teslas) and
$\mathbf{H}_{ext}$ is the externally applied magnetic field. The full
energy landscape then becomes $U(\mathbf{m})=U_K +U_P+ U_H$ and reads:

\begin{equation}
U(\mathbf{m})=K\left[D(\mathbf{\hat{n}}_D\cdot\mathbf{m})^2-(\mathbf{\hat{n}}_K\cdot\mathbf{m})^2-2\mathbf{h}\cdot\mathbf{m}\right],
\end{equation}
where $D\equiv(K_P/K)$, $\mathbf{h}=\mathbf{H}_{ext}/H_K$ and
$\mathbf{\hat{n}}_K$ is the unit vector pointing in the orientation of
the uniaxial anisotropy axis. Such an energy landscape has stable
magnetic configurations parallel and anti-parallel to
$\mathbf{\hat{n}}_K$. The effective interaction field
$\mathbf{H}_{\mathrm{eff}}$ is then given by

\begin{equation}
\mathbf{H}_{\mathrm{eff}}=-\frac{1}{M_S V}\nabla_{\mathbf{m}}U(\mathbf{m})=-H_K\left[D(\mathbf{\hat{n}}_D\cdot\mathbf{m})\mathbf{\hat{n}}_D-(\mathbf{\hat{n}}_K\cdot\mathbf{m})\mathbf{\hat{n}}_K-\mathbf{h}\right],
\end{equation}
and the deterministic LLG dynamics can then be expressed as:

\bea
\label{eq:dot}
\mathbf{\dot{m}}&=&-\mathbf{m}\times\left[(\mathbf{\hat{n}}_K\cdot\mathbf{m})\mathbf{\hat{n}}_K-D(\mathbf{\hat{n}}_D\cdot\mathbf{m})\mathbf{\hat{n}}_D+\mathbf{h}\right]\nonumber\\
&-&\alpha\mathbf{m}\times\left[\mathbf{m}\times\left((\mathbf{\hat{n}}_K\cdot\mathbf{m})\mathbf{\hat{n}}_K-D(\mathbf{\hat{n}}_D\cdot\mathbf{m})\mathbf{\hat{n}}_D+\mathbf{h}\right)\right]\nonumber\\
&-&\alpha I\mathbf{m}\times\left(\mathbf{m}\times\mathbf{\hat{k}}\right)+\alpha^2 I\mathbf{m}\times\mathbf{\hat{k}},
\eea
where we have defined $I=j/(\alpha H_K)$, introduced the natural timescale $\tau=\gamma' H_K t$, and have chosen our z-axis to be aligned with the spin polarization axis ($\hat{\mathbf{k}}\equiv\hat{\mathbf{n}}_p$).

\begin{figure}
	\begin{center}
	\centerline{\includegraphics [clip,width=4in, angle=0]{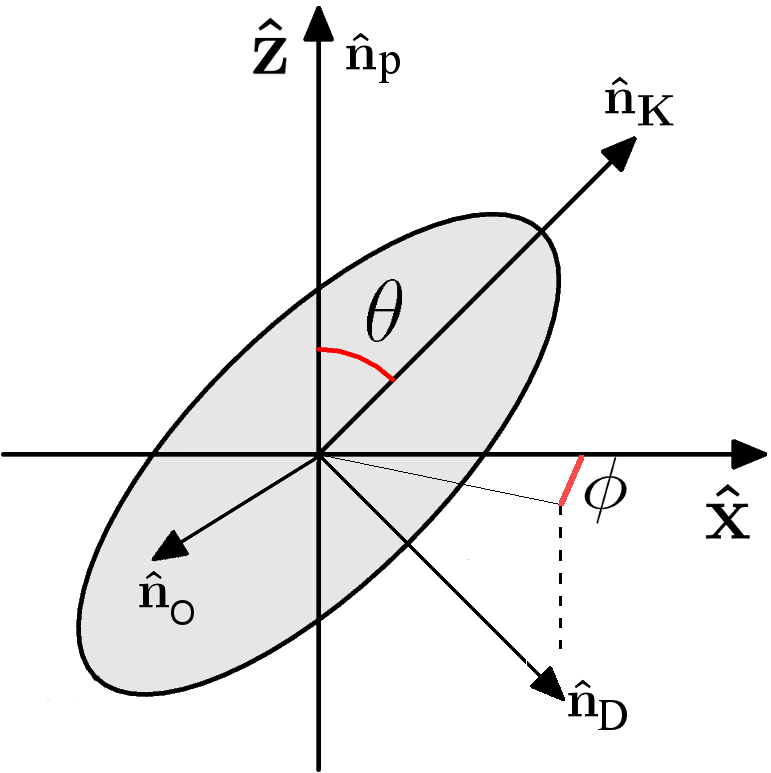}}
	\caption{Uniaxial and hard-axis anisotropy axes tilted with respect to polarizer by angle $\theta$.}
	\label{F1}
	\end{center}
\end{figure} 

In practice, the uniaxial and hard-axis axes cannot simply be oriented
in any given direction but are, in fact, perpendicular~(see
Fig.~1). We can write the equations of motion for the magnetization in
a rotated coordinate system, where $\mathbf{\hat{n}}_K$,
$\mathbf{\hat{n}}_D$ and $\mathbf{\hat{n}}_O$ define our new
coordinate system. Let $\mathbf{\hat{n}}_K=(n_x,0,n_z)$,
$\mathbf{\hat{n}}_D=(n_z\sqrt{1-n^2},n,-n_x\sqrt{1-n^2})$ and
$\mathbf{\hat{n}}_O=(-n_zn,\sqrt{1-n^2},n_xn)$ be the orientation of
the uniaxial and hard-axis anisotropy vectors with respect to our
original coordinate system (where $n_z=\cos\theta$, $n_x=\sin\theta$
and $n = \sin\phi$). We can then write the dynamical equations in
terms of the projections $q\equiv\mathbf{\hat{n}}_K\cdot\mathbf{m}$,
$s\equiv\mathbf{\hat{n}}_D\cdot\mathbf{m}$ and
$p\equiv\mathbf{\hat{n}}_O\cdot\mathbf{m}$ of the magnetization vector
along this new set of axes. Written componentwise, they are:

\begin{eqnarray}
\label{eq:detdyn}
\dot{s}&=&qp-\alpha\left[(In_x\sqrt{1-n^2}+Ds)(1-s^2)+qs(In_z+q)+In_xnsp\right]\\
\dot{q}&=&Dsp+\alpha\left[(In_z+q)(1-q^2)+Iqn_x(s\sqrt{1-n^2}-np)+Ds^2q\right]\\
\dot{p}&=&-(D+1)qs+\alpha\left[sp(In_x\sqrt{1-n^2}+Ds)-qp(In_z+q)+In_xn(1-p^2)\right].
\end{eqnarray}  

\section{Thermal Effects}

Thermal effects are included by considering uncorrelated fluctuations
in the effective interaction field:
$\mathbf{H}_{\mathrm{eff}}\rightarrow\mathbf{H}_{\mathrm{eff}}+\mathbf{H}_{th}$. These
transform the LLG equation into its Langevin form. We model the
stochastic contribution $\mathbf{H}_{th}$ by specifying its
correlation properties, namely:

\begin{eqnarray}
\label{eq:corrprop}
&\langle \mathbf{H}_{th}\rangle=0 \\
&\langle H_{th,i}(t)H_{th,k}(t')\rangle=2C\delta_{i,k}\delta(t-t')
\end{eqnarray} 
The effect of the random torque~$\mathbf{H}_{th}$ is to produce a diffusive
random walk on the surface of the $\mathbf{M}$-sphere. An associated
Fokker-Planck equation describing such dynamics was constructed by
Brown~\cite{Brown}. At long times, the system attains thermal equilibrium;
by the fluctuation-dissipation theorem we have

\begin{equation}
C=\frac{\alpha k_BT}{2K(1+\alpha^2)}=\frac{\alpha}{2(1+\alpha^2)\xi},
\end{equation}
where $\xi = K/k_BT$ is the barrier height due to the uniaxial anisotropy energy divided by the thermal energy.

Setting the external magnetic field to zero and considering only thermal
fluctuations, the stochastic LLG equation reads:

\begin{equation}
\label{eq:Langevindynamics}
\dot{m}_i=A_i(\mathbf{m})+B_{ik}(\mathbf{m})\circ H_{th,k}
\end{equation}
where

\begin{eqnarray}
\mathbf{A}(\mathbf{m})&=&\alpha I\left[\alpha\mathbf{m}\times\mathbf{\hat{k}}-\mathbf{m}\times\left(\mathbf{m}\times\mathbf{\hat{k}}\right)\right]\nonumber\\
&-&(\mathbf{\hat{n}}_K\cdot\mathbf{m})\left[\mathbf{m}\times\mathbf{\hat{n}}_K-\alpha\left(\mathbf{\hat{n}}_K-(\mathbf{\hat{n}}_K\cdot\mathbf{m})\mathbf{m}\right)\right]\nonumber\\
&+&D(\mathbf{\hat{n}}_D\cdot\mathbf{m})\left[\mathbf{m}\times\mathbf{\hat{n}}_D-\alpha\left(\mathbf{\hat{n}}_D-(\mathbf{\hat{n}}_D\cdot\mathbf{m})\mathbf{m}\right)\right],\\
B_{ik}(\mathbf{m})&=&-\epsilon_{ijk}m_j-\alpha(m_i m_k - \delta_{ik}).
\end{eqnarray}
and `$\circ H_{th,k}$' means that the multiplicative noise term in
the dynamical equation is to be interpreted in the sense of Stratonovich calculus~\footnote{In modeling thermal effects through stochastic
  contributions in a system like ours, we are interested in considering the
  white noise limit of a potentially colred noise
  process~\cite{WongZakai}. Stratonovich calculus is then to be preferred
  over \^Ito calculus as ascertained by the Wong-Zakai theorem.}~\cite{Karatsas}.

We numerically solve the above Langevin equations by using a standard
second-order Heun scheme to ensure proper convergence. At each time
step, the strength of the random kicks is given by the
fluctuation-dissipation theorem. Statistics were gathered from an
ensemble of~$5000$ events with a natural integration stepsize
of~$0.01$. For concreteness, we set the damping constant~$\alpha=0.04$
and the barrier height to~$\xi=80$. To explore the simulations for
long time regimes, the necessary events were simulated in parallel on
an NVidia Tesla C2050 graphics card. To generate the large number of
necessary random numbers, we chose a proven combination~\cite{Nguyen}
of the three-component combined Tausworthe ``taus88"~\cite{Ecuyer} and
the 32-bit ``Quick and Dirty" LCG~\cite{Press}. The hybrid generator
provides an overall period of around $2^{121}$.


\section{Switching Dynamics}

In experiments, one is generally interested in understanding how the
interplay between thermal noise and spin torque effects switch an
initial magnetic orientation from parallel to antiparallel.  The role
of spin torque from an energy landscape point of view can be
elucidated by considering how energy is pumped into the system. As in
the previous section, the monodomain's magnetic energy (for
$\mathbf{h}=0$) can be written in dimensionless form:

\begin{equation}
\label{eq:eps}
\epsilon(\mathbf{m})\equiv U(\mathbf{m})/K=D(\mathbf{\hat{n}}_D\cdot\mathbf{m})^2-(\mathbf{\hat{n}}_K\cdot\mathbf{m})^2=Ds^2-q^2.
\end{equation}
Stable states with minimum energy have $q=\pm1$ and thus
$\epsilon=-1$.  The change in energy over time can then be written as:
\begin{equation}
\label{eq:dotenergy}
\dot{\epsilon}=2\left[Ds\dot{s}-q\dot{q}\right].
\end{equation}
The dynamical equations~(\ref{eq:Langevindynamics}) for $\dot{s}$
and $\dot{q}$ can be plugged back in to~(\ref{eq:eps}) to obtain a
dynamical equation for the energy:
\begin{equation}
\label{eq:energy}
\dot{\epsilon}=f(s,q)+g(s,q)\circ\dot{W}\, .
\end{equation}
Nanomagnets typically operate in a regime where the nonconservative parts
of the dynamics, including the effects of damping, STT and thermal noise,
act on timescales that are long compared to the energy-conserving piece of
the Hamiltonian. Trajectories are therefore expected to remain close to
periodic Hamiltonian orbits for a long time, slowly drifting from one orbit
to another. Using this information, we average the energy evolution
equation~(\ref{eq:energy}) over constant energy trajectories, thereby
obtaining a diffusion equation in energy
space~\cite{Apalkov,Katie,Serpico}. From~(\ref{eq:dot}) we have

\begin{eqnarray}
\label{eq:dot0}
\dot{s}^0&=&-q^0p^0\\
\dot{q}^0&=&-Ds^0p^0\\
\dot{p}^0&=&(D+1)q^0s^0,
\end{eqnarray}
the solutions of which can be expressed in terms of Jacobi elliptic
functions:
\begin{eqnarray}
s^0(t)&=&\mp \sqrt{D\frac{1+\epsilon}{D+1}}\cn\left[\sqrt{D-\epsilon}t,k^2\right]\\
q^0(t)&=&\pm \sqrt{\frac{D-\epsilon}{D+1}}\dn\left[\sqrt{D-\epsilon}t,k^2\right]\\
p^0(t)&=&-\sqrt{1+\epsilon}\sn\left[\sqrt{D-\epsilon}t,k^2\right],
\end{eqnarray}
where $k^2\equiv D\frac{1+\epsilon}{D-\epsilon}$. The period of these
trajectories as a function of energy can be expressed as a complete
elliptic integral of the first kind:
\be
\label{eq:period_T}
T(\epsilon)=\frac{4}{\sqrt{D-\epsilon}}\int_0^1\frac{dx}{\sqrt{(1-x^2)(1-k^2x^2)}}=\frac{4}{\sqrt{D-\epsilon}}\mathrm{K}(k^2).
\ee 
A sample of these trajectories for energies greater and less than
$0$ are shown in Fig.~2. The $\epsilon<0$ dynamics consist of the
magnetization precessing around the uniaxial axis
($\mathbf{\hat{n}}_K$), thus remaining in the energy minimum's basin of
attraction.  On the other hand, for $\epsilon>0$ the magnetization
precesses around the hard axis ($\mathbf{\hat{n}}_D$). Magnetic
switching, defined as a transition from one $\epsilon<0$ basin to
another, must proceed through an intermediate $\epsilon>0$. As a
result, we will consider $\epsilon=0$ to be the critical threshold
energy that a monodomain must surpass to achieve switching.

\begin{figure}
	\begin{center}
	\centerline{\includegraphics [clip,width=4in, angle=0]{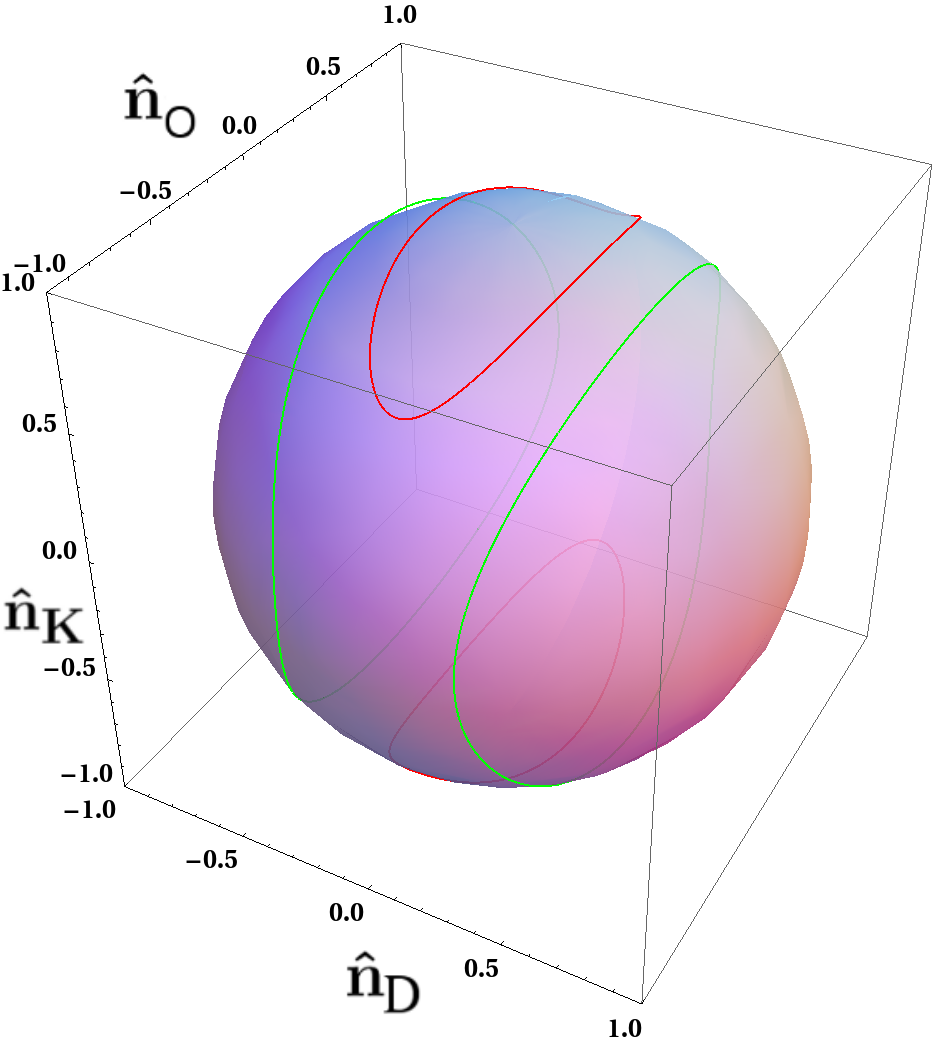}}
	\caption{Constant enegrgy trajectories. $\epsilon<0$ trajectories are shown in red whereas $\epsilon>0$ trajectories are shown in green.}
	\label{F2}
	\end{center}
\end{figure} 

The qualitative properties of such trajectories are better understood
by parametrizing them geometrically as follows. The energy
landscape~(\ref{eq:eps}) determines the geometrical conditions these
trajectories must satisfy. For $\epsilon>0$ we have
\bea
&\frac{D}{\epsilon}s^2-\frac{1}{\epsilon}q^2=1\\
&s^2+q^2+p^2=1.
\eea
These are satisfied by the parametrization:
\bea
s^0(w)&=&\pm \sqrt{\frac{|\epsilon|}{D}}\cosh(w)\\
q^0(w)&=&\sqrt{|\epsilon|}\sinh(w)\\
p^0(w)&=&\pm\sqrt{1+|\epsilon|}\sqrt{1-\gamma_+^2\cosh^2(w)},\\
\gamma_+^2&=&\frac{\epsilon(D+1)}{D(\epsilon+1)}.
\eea
where the limits of validity of $-\acosh(1/\gamma_+)<w<\acosh(1/\gamma_+)$
are found by imposing the extra condition that for $p=0$, $s^2+q^2=1$. 
The identical reasoning carried out for $\epsilon<0$ leads to the
parametrization:
\bea
\label{eq:geoparam}
s^0(w)&=&\pm \sqrt{\frac{|\epsilon|}{D}}\sinh(w)\\
q^0(w)&=&\pm\sqrt{|\epsilon|}\cosh(w)\\
p^0(w)&=&\pm\sqrt{\frac{D+|\epsilon|}{D}}\sqrt{1-\gamma_-^2\cosh^2(w)},\\
\gamma_-^2&=&\frac{|\epsilon|(D+1)}{D+|\epsilon|},
\eea
with limits $-\acosh(1/\gamma_-)<w<\acosh(1/\gamma_-)$.

We are now ready to average~(\ref{eq:dotenergy}) over constant energy
orbits. The averaging procedure is simplified by noting that we are
interested in switching behavior starting from the $\epsilon<0$ basin
and that in that basin the constant energy trajectories are symmetric
with respect to the $\mathbf{\hat{n}}_K$ axis. As such, the majority
of terms obtained by inserting~(\ref{eq:detdyn})
into~(\ref{eq:dotenergy}) will average to zero. The remaining nonzero
terms lead to the averaged energy equation:
\bea
\label{eq:epsdot}
\langle\partial_{\mathrm{t}}\epsilon\rangle&=&-2\alpha\left[In_z(1+\epsilon)\langle q\rangle+(1+\epsilon)\langle q^2\rangle+D(D-\epsilon)\langle s^2\rangle\right]\nonumber\\
&+&2\sqrt{\frac{\alpha(D+1)}{\xi}}\sqrt{\langle q^2\rangle+\frac{D-\epsilon}{D+1}\epsilon}\circ\dot{W} 
\eea 
where $\langle\cdot\rangle$ implies averaging over constant energy
orbits as described.  Construction of the stochastic term requires
averaging the variance of~(\ref{eq:dotenergy}) following the rules of
additivity for Gaussian random variables.  It is seen
from~(\ref{eq:epsdot}) that the applied current factors into this
equation only in the form $In_z$, where $n_z$ is the cosine of the
angular tilt between uniaxial anisotropy and polarizer axes. This
leads to a more general form of a result obtained in our previous
work~\cite{APL}, namely that scalings between switching times and
applied current will be functionally identical independent of the
incoming angle of the spin-polarized current.  The only caveat here is
that every current must be rescaled by multiplying it by~$n_z$.
Analogously, the azimuthal tilt parameter~$n$ does not appear at all
in the energy-averaged equation, implying that only coplanar setups
between polarizer, easy and hard axes need to be considered in what
follows.

With this understanding, we can now omit the notation~$n_z$ and
consider macrospin setups with no angular tilt between polarizer and
uniaxial anisotropy axes. The next step is to show how the necessary
averages can be computed explicitly. As an example, consider $\langle
q\rangle$; its average is given by:

\be 
\langle q\rangle=\frac{1}{T(\epsilon)}\int_0^Tdtq^0(t),
\ee
where the integration is over time and one uses the expression for
$q^0$ in terms of its Jacobi elliptic function. We can express the same
integral in terms of the geometrical parametrization~(\ref{eq:geoparam}). 
In fact
\be
\langle q\rangle=\frac{1}{T(\epsilon)}\int_0^Tdtq^0(t)=\frac{1}{T(\epsilon)}\oint dw \frac{ds^0/dw}{\dot{s}^0}q^0(w).
\ee
Putting together the expression for $\dot{s}^0$ from~(\ref{eq:dot0})
and, again, employing the geometrical parametrizations
from~(\ref{eq:geoparam}), we obtain
\be
\langle q\rangle=-\frac{4}{T(\epsilon)\sqrt{D+1}}\int_0^{\acosh(1/\gamma)}dw\frac{\cosh(w)}{\sqrt{1-\gamma^2\cosh^2(w)}}=-\frac{2\pi}{T(\epsilon)\sqrt{D+1}},
\ee
where, for notational convenience, we redefine $\gamma=\gamma_-$. 

Repeating the procedure for the other terms in (8), we write
\bea
\langle q^2\rangle(\epsilon)&=&\frac{4}{T(\epsilon)\sqrt{D+1}}\sqrt{\frac{D}{D+1-\gamma^2}}\eta_1\\
\langle s^2\rangle(\epsilon)&=&\frac{4}{T(\epsilon)\sqrt{D+1}}\sqrt{\frac{D}{D+1-\gamma^2}}\frac{1}{D}\left(\eta_1-\gamma^2\eta_0\right),
\eea
with
\bea
\eta_0(\epsilon)&=&\int_0^1\frac{dx}{\sqrt{(\gamma^2+(1-\gamma^2)x^2)(1-x^2)}}=\frac{\mathrm{K}(1-\frac{1}{\gamma^2})}{\gamma}=\mathrm{K}(1-\gamma^2)\\
\eta_1(\epsilon)&=&\int_0^1dx\sqrt{\frac{\gamma^2+(1-\gamma^2)x^2}{1-x^2}}=\gamma\mathrm{E}(1-\frac{1}{\gamma^2})=\mathrm{E}(1-\gamma^2)\label{eq:Period}\\
T(\epsilon)&=&4\sqrt{\frac{D+1-\gamma^2}{D(D+1)}}\mathrm{K}(1-\gamma^2)=4\sqrt{\frac{D+1-\gamma^2}{D(D+1)}}\eta_0(\gamma)\\
\gamma(\epsilon)&=&\frac{|\epsilon|(D+1)}{D+|\epsilon|},
\eea
where $\mathrm{K}(x)$ and $\mathrm{E}(x)$ are elliptic functions of
the first and second kind respectively~(Appendix~A). The energy
equation for the dynamics starting in the negative energy well
($0<|\epsilon|<1$) then reads (expressed in terms of
$\gamma$\footnote{We can allow ourselves the freedom to switch between
  expressions involving $\gamma$ and
  $\epsilon$. $\gamma^2=\frac{|\epsilon|(D+1)}{D+|\epsilon|}$ is a
  monotonically increasing function of $|\epsilon|$ with the
  convenient property that $|\epsilon|=1\to\gamma=1$ and
  $|\epsilon|=1\to\gamma=1$. As such limits written in terms of
  $\gamma$ and $\epsilon$ are equivalent.}):

\bea
\label{eq:energydiff}
\partial_{\mathrm{t}}|\epsilon|(\gamma)&=&\frac{8\alpha}{T(\gamma)}\left(\frac{D(D+1)^{1/3}}{D+1-\gamma^2}\right)^{3/2}\left[\eta_1(\gamma)-\gamma^2\eta_0(\gamma)+\frac{1-\gamma^2}{D}\left(\eta_1(\gamma)-\frac{\pi I}{2}\sqrt{\frac{D+1-\gamma^2}{D}}\right)\right]\nonumber\\
&+&4\sqrt{\frac{\alpha}{\xi T(\gamma)}}\left(\frac{D(D+1)}{D+1-\gamma^2}\right)^{1/4}\sqrt{\eta_1(\gamma)-\frac{D\gamma^2}{D+1-\gamma^2}\eta_0(\gamma)}\circ\dot{W}.
\eea

In outlining the steps above, we have succeded in reducing the
multidimensional complexity of the full magnetization dynamics to a
one-dimensional stochastic differential equation whose properties we
now proceed to study. The relation to previous studies~\cite{APL,PRB}
on uniaxial macrospins can be rederived by considering
equations~(\ref{eq:dot0}) and~(\ref{eq:energydiff}) in the limit
$D\to0$. Doing so leads to the much simplified energy diffusion
equation~\cite{MMM}
\be
\partial_{\mathrm{t}}|\epsilon|=2\alpha\sqrt{|\epsilon|}(1-|\epsilon|)(\sqrt{|\epsilon|}-I)+2\sqrt{\frac{\alpha}{\xi}|\epsilon|(1-|\epsilon|)}\circ\dot{W},
\ee 
with a stable energy minimum at $\epsilon=-1$ and saddle point at
$\epsilon=-I^2$. This equation also shows that, for currents
$I>I_c=\max_{|\epsilon|\in[0,1]}\sqrt{|\epsilon|}=1$, the flow becomes negative
for all values of $|\epsilon|$, so that in this regime all states will
deterministically switch.

\section{Stability Analysis}

In the absence of applied currents,
$\partial_{\mathrm{t}}|\epsilon|>0$ as determined
by~(\ref{eq:energydiff}) and $\epsilon$ flows toward its minimum value
of~$-1$ (the stable fixed point of its dynamics). To understand
switching one must therefore investigate under what circumstances
$\epsilon$ may become greater than zero, thus implying a transition
from the red to the green trajectories discussed in Fig.~2.  It
suffices to understand the behavior of the energy flow at the stable
fixed point of the zero current dynamics ($|\epsilon|=\gamma=1$) and
at the energy threshold for switching ($|\epsilon|=\gamma=0$).  These
limiting flows are found by computing the integrals $\eta_0(\gamma)$
and $\eta_1(\gamma)$. One finds:

\bea
&\lim\limits_{\gamma\to1}\eta_0(\gamma)\sim\frac{\pi}{2}+\frac{\pi}{8}\frac{D}{D+1-\gamma^2}(1-\gamma^2)+o((1-\gamma^2)^2)\\
&\lim\limits_{\gamma\to1}\eta_1(\gamma)\sim\frac{\pi}{2}-\frac{\pi}{8}\frac{D}{D+1-\gamma^2}(1-\gamma^2)+o((1-\gamma^2)^2),
\eea
and

\bea
&\lim\limits_{\gamma\to0}\eta_0(\gamma)\sim\log(\gamma)+o(\log(\gamma)\gamma^2)\label{eq:etalim}\\
&\lim\limits_{\gamma\to0}\eta_1(\gamma)\sim 1+o(\gamma).
\eea
Using these results in~(\ref{eq:energydiff}), one finds that in the absence
of thermal noise (i.e., the zero temperature limit), the limiting flows are:
\bea
\label{eq:energyflowat0}
&\lim\limits_{\gamma\to0}\partial_{\mathrm{t}}|\epsilon|=\frac{8\alpha}{T}\frac{\sqrt{D}}{D+1}\left[D+1-\frac{\pi I}{2}\sqrt{\frac{D+1}{D}}\right]\\
\label{eq:energyflowat1}
&\lim\limits_{\gamma\to1}\partial_{\mathrm{t}}|\epsilon|=\frac{4\pi\alpha}{T}\frac{\sqrt{D+1}}{D}(1-\gamma^2)\left[D+2-2I\right].
\eea

As expected, both eqs.~(\ref{eq:energyflowat0})
and~(\ref{eq:energyflowat1}) show how the qualitative behavior of the
deterministic energy flow can be tuned by the value of applied
current. The stability of the zero current stable fixed point
($|\epsilon|=\gamma=1$) can in fact be be rendered unstable once
(\ref{eq:energyflowat1}) changes sign at the applied current value:
\be 
I_c^{|\epsilon|=1}\equiv I_C^1=\frac{D+2}{2}. 
\ee
Analogously, at $|\epsilon|=\gamma=0$ the flow is either positive or negative
depending on the applied current. Using~(\ref{eq:energyflowat0}), we find
that the sign of the flow switches from positive to negative at 
\be
\label{eq:IC0}
I_c^{|\epsilon|=0}\equiv I_C^0=\frac{2}{\pi}\sqrt{D(D+1)}.  
\ee 
Identical expressions for the critical currents have recently been derived
by Taniguchi~\cite{TaniguchiE} via different means. These two critical
stability thresholds are identical at a critical value of $D$:

\be
D_0=2\frac{\pi^2-4}{16-\pi^2}\left[1+2\frac{\sqrt{4+2\pi^2}}{\pi^2-4}\right]\sim5.09.
\ee

The value of $D$ relative to $D_0$, as well as the applied current, will select
between two qualitatively different dynamical regimes. We now proceed to
explore both.

\subsection{$D>D_0$}

When the ratio between hard and easy axis anisotropies is greater than the
threshold value $D_0$, $I_C^0>I_C^1$ (see Fig.~3). Unless the applied current is greater
than $I_C^0$, deterministic switching of the monodomain cannot be achieved.
This differs from conclusions drawn by previous work~\cite{Sun} where the
critical current for deterministic switching was assumed to be that which
renders the stable fixed point ($|\epsilon|=\gamma=1$) unstable.

For small currents such that $I<I_C^1$, on the other hand, the
deterministic flow pushes the energy toward the zero current stable fixed
point. In such a scenario, switching can only occur via thermal activation
over the $|\epsilon|=\gamma=0$ effective barrier.

Finally, for currents $I_C^0>I>I_C^1$, the zero current stable fixed
point has been rendered unstable while the flow at the switching
energy threshold remains positive. The monodomain will move
(deterministically) from the $|\epsilon|=\gamma=1$ fixed point, but
switching over the $|\epsilon|=\gamma=0$ must still take place via
thermal activation. This implies that a new stable energy equilibrium
exists for some value $0<\tilde{\gamma}_S<1$. The precessional
macrospin dynamics will then manifest itself in the form of stable
limit cycles. This has been observed both
experimentally~\cite{Krivorotov} and numerically~\cite{Katie}. These
three dynamical scenarios are displayed in Fig.~4 where the
deterministic energy flow is plotted as a function of energy.

\begin{figure}
	\begin{center}
	\centerline{\includegraphics [clip,width=5in, angle=0]{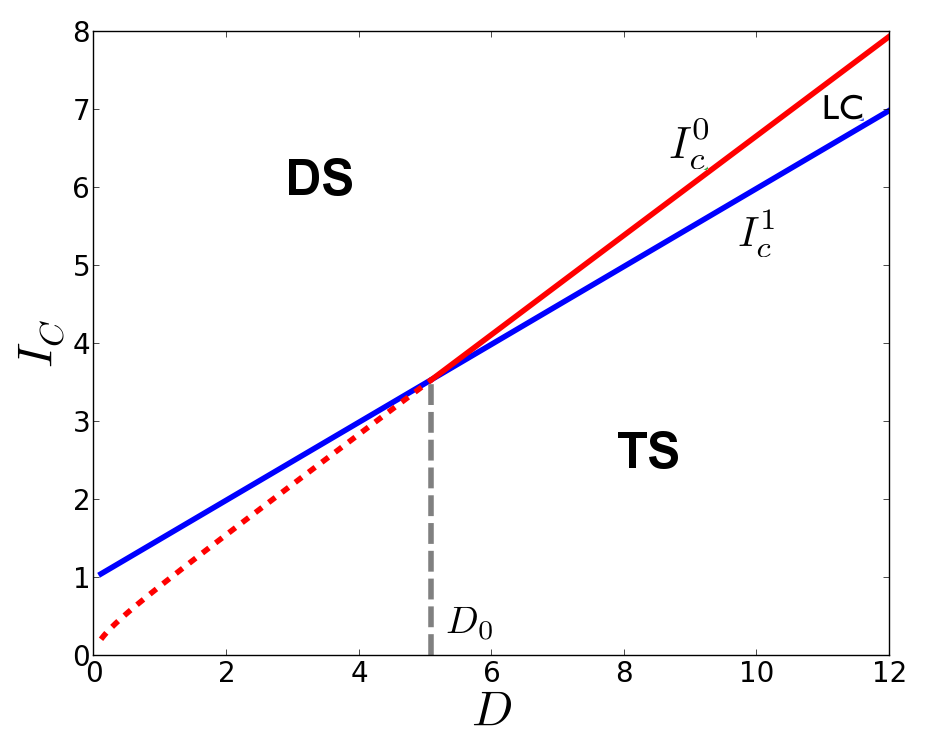}}
	\caption{Critical currents versus the ratio of the hard and easy axis anisotropies $D$. The blue curve is $I_C^1$ and the red curve
is $I_C^0$. For $D<D_0$, currents greater than $I_C^1$ lead to deterministic switching (labeled DS). For $D>D_0$ currents between $I_C^0$ and $I_C^1$ lead to 
limit cycles (LC).  Limit cycles can also occur for currents just below and approximately equal to $I_C^1$ for $D<D_0$, as shown in Fig. 7}
	\label{F3}
	\end{center}
\end{figure}

\begin{figure}
	\begin{center}
	\centerline{\includegraphics [clip,width=5in, angle=0]{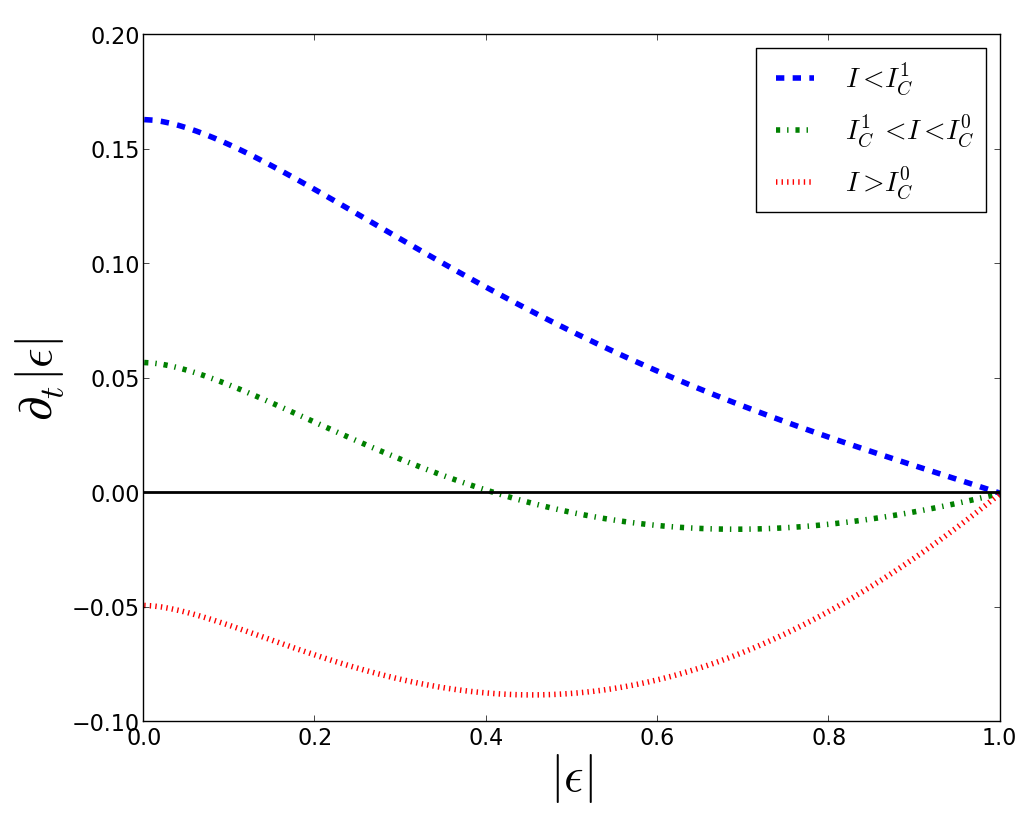}}
	\caption{Three sample regimes of deterministic energy flow $\dot{\epsilon}$ as a function of energy for $D>D_0$. Coloring (online) is included to better distinguish the various curves. a) $I<I_C^1$: Subcritical regime, thermal noise must oppose a positive energy flow to achieve switching; b) $I_C^1<I<I_C^0$: Limit cycle regime; and c) $I>I_C^0$: Supercritical regime, negative flow leads to determinist switching.}
	\label{F4}
	\end{center}
\end{figure}

\subsection{$D<D_0$}

When the ratio between hard and easy axis anisotropies is smaller than
the threshold value~$D_0$, we have $I_C^1>I_C^0$ (see Fig 3). The
threshold critical current that switches all magnetic states
deterministically is now $I_C^1$~\cite{Sun}. Above this value of the
applied current, switching will occur independently of thermal noise
effects.

For applied currents such that $I_C^0<I<I_C^1$, a saddle point emerges
for some value $0<\tilde{\gamma}_U<1$. For switching to occur, thermal
activation must move the monodomain energy past this saddle point.  As
the current is lowered, $\tilde{\gamma}_U$ becomes progressively
smaller until the limiting value $\tilde{\gamma}_U=0$. At this point,
switching requires thermal activation throughout the whole energy
range. These three dynamical scenarios are displayed in Fig.~5 where
the deterministic energy flow is plotted as a function of energy.

It is interesting to note that the condition $D<D_0$ can also result in the
appearance of limit cycles. Limit cycle regimes in fact can be seen for
very small applied ranges of current less than $I_C^0$~(Fig.~6).

The uniaxial macropin model is a particular case of a $D<D_0$ model
($D=0$). By taking the limit $D\to0$, we find $I_C^0\to0$ and $I_C^1\to1$,
as found in previous work.~\cite{APL} For all values of the applied
current strictly between~0 and~1, uniaxial macrospin switching must take
place via thermal activation over an effective energy barrier.

\begin{figure}
	\begin{center}
	\centerline{\includegraphics [clip,width=5in, angle=0]{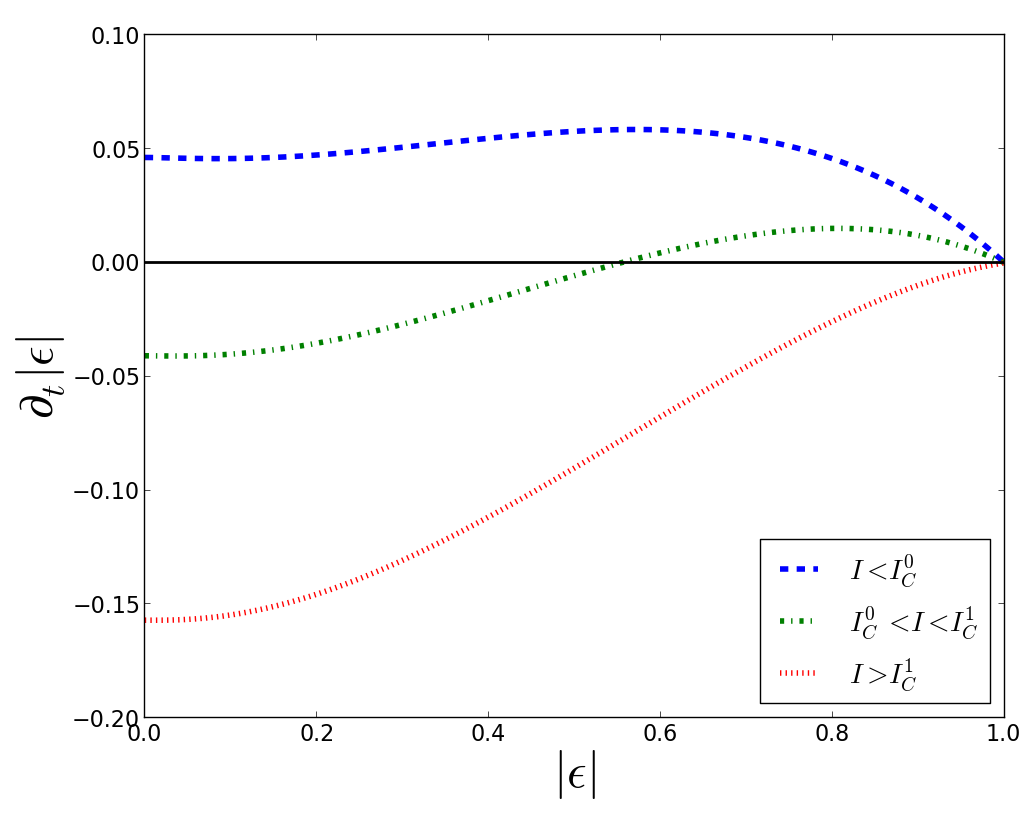}}
	\caption{Three sample regimes of deterministic energy flow $\dot{\epsilon}$ as a function of energy for $D<D_0$. Coloring (online) is included to better distinguish the various curves. a) $I<I_C^0$: Subcritical regime, thermal noise must oppose a positive energy flow to achieve switching; b) $I_C^0<I<I_C^1$: Crossover regime, switching is still achieved via thermal activation but the unstable equlibirum has now shifted; and c) $I>I_C^1$: Supercritical regime, negative flow leads to determinist switching.}
	\label{F5}
	\end{center}
\end{figure}

\begin{figure}
	\begin{center}
	\centerline{\includegraphics [clip,width=5in, angle=0]{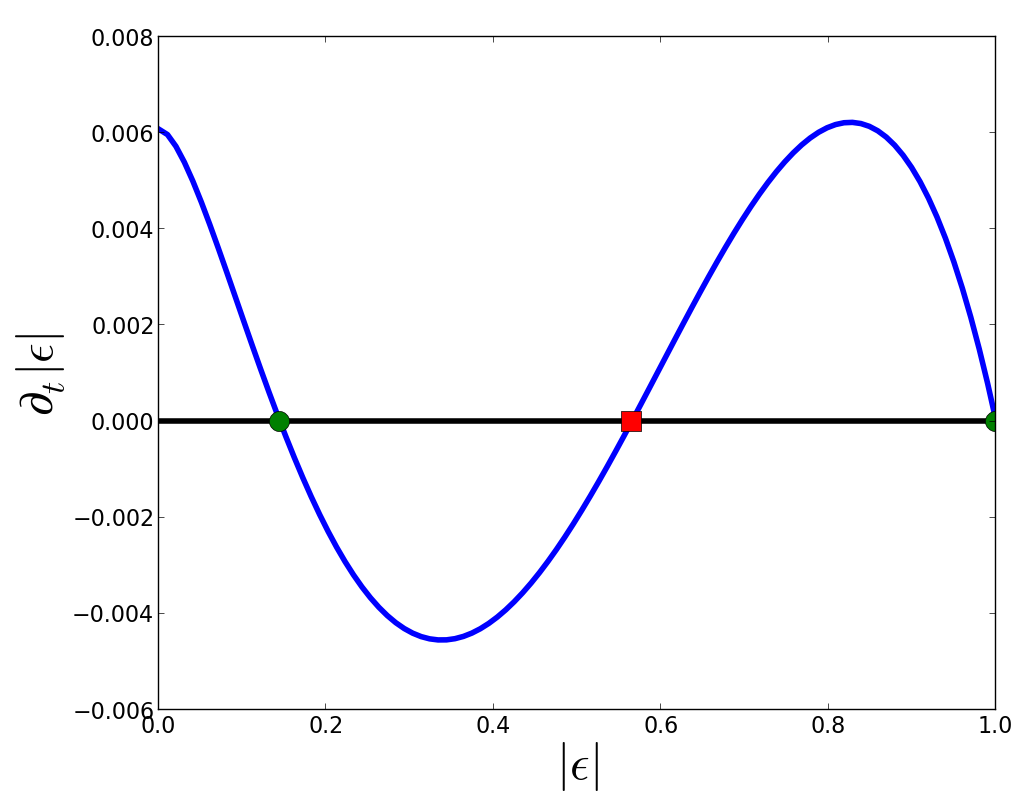}}
	\caption{Energy flow for a $D=4$ macrospin and applied current $I=2.82<I_C^0$. Circles and squares respectively represent stable and unstable equilibria. For these parameters ($D=4$ and $I=2.82$) , two stable equilibria of the zero temperature dynamics coexist.}
	\label{F6}
	\end{center}
\end{figure}

\section{Limits of the CEOA Approach}

In this section we will determine under what conditions the CEOA
approach is valid.  The fundamental assumption is that the
deterministic precession timescale of the constant energy orbits be
small compared to the timescale over which significant energy
diffusion due to noise and spin-torque transfer takes place. We now
quantify this condition and obtain precise limits in terms of the
applied current intensity $I$ and the anisotropy ratio $D$.  For our
approximations to be valid, the averaged energy flow
($T(\epsilon)\partial_t|\epsilon|$) over any given orbit must be small
compared to the maximum allowable energy ($0<|\epsilon|<1$):
\be
\label{eq:validity}
\underaccent{\epsilon}{\mathrm{max}}\: T(\epsilon)\lvert\partial_{\mathrm{t}}|\epsilon|\rvert=\underaccent{\gamma}{\mathrm{max}}\: T(\gamma)\lvert\partial_{\mathrm{t}}|\epsilon|\rvert\ll1,
\ee
where we continue to use the variable
$\gamma(\epsilon)=|\epsilon|(D+1)/(D+|\epsilon|)$. To proceed, we
consider the deterministic drift and random components of the energy
flow separately. Noise averaging~(\ref{eq:energydiff}) will give the
contribution of the drift to the energy flow. The intensity of the
orbit averaged drift then reads:
\be
\label{eq:orbavdrift}
T(\gamma)\lvert\langle\partial_{\mathrm{t}}|\epsilon|\rangle\rvert=8\alpha\left(\frac{D(D+1)^{1/3}}{D+1-\gamma^2}\right)^{3/2}\left\lvert\eta_1(\gamma)-\gamma^2\eta_0(\gamma)+\frac{1-\gamma^2}{D}\left(\eta_1(\gamma)-\frac{\pi I}{2}\sqrt{\frac{D+1-\gamma^2}{D}}\right)\right\rvert,
\ee
where angular brackets denote noise averaging. This expression can be
shown to be finite for all values of $\gamma$, $I$ and
$D$. Furthermore, for applied currents greater than
$I_C^0$,~(\ref{eq:orbavdrift}) has a maximum at $\gamma=0$; its value
in this limit is given by~(\ref{eq:energyflowat0}). Enforcing the
conditions for validity discussed above results in an upper bound for
the current:
\be
\label{eq:IM}
I\ll (1+\frac{1}{8\alpha\sqrt{D}})I_C^0\equiv I_M.
\ee
In the limit $D\to 0$, this expression converges to the correct
uniaxial limit inequality $I\ll (4\pi\alpha)^{-1}$, discussed
previously in~\cite{MMM}.

When switching occurs deterministically (and $I>I_C^0$), this bound is
the sole limit to the validity of the CEOA approximation. This is
because random contributions to the energy flow dynamics have zero
mean. However, in scenarios where switching occurs due to thermal
activation, one must consider the standard deviation of random
contributions to the energy flow. These determine the rate of the
rare events that drive the system over its confining energy
barrier. The energy flow's standard deviation is
$\sqrt{\langle[\partial_{\mathrm{t}}|\epsilon|-\langle\partial_{\mathrm{t}}|\epsilon|\rangle]^2\rangle}$. The
orbit-averaged rate of random events driving noise-induced energy
diffusion can then be written as:
\be
\label{eq:orbavnoise}
T(\gamma)\sqrt{\langle[\partial_{\mathrm{t}}|\epsilon|-\langle\partial_{\mathrm{t}}|\epsilon|\rangle]^2\rangle}=8\sqrt{\frac{\alpha}{\xi}\eta_0(\gamma)\left(\eta_1(\gamma)-\frac{D\gamma^2}{D+1-\gamma^2}\eta_0(\gamma)\right)},
\ee
which can be shown to be both a monotonically decreasing function of
$\gamma$ and logarithmically divergent for $\gamma\to 0$. The latter
is due to the limiting behavior of $\eta_0(\gamma)$
(see~(\ref{eq:etalim})) for all $D>0$. The physical origin of this
behavior lies in the divergence of the constant energy precessional
period as the zero energy switching threshold is reached (a fact
also pointed out by Taniguchi et al.~\cite{TaniguchiE}).

The relevant maxima of~(\ref{eq:orbavnoise}) will be located at the
energy flow's saddle point $\gamma_U$. Unfortunately, solving for
$\gamma_U$ while imposing that the RHS of~(\ref{eq:orbavnoise}) be
much less than $1$ results in a set of transcendental equations which
can only be solved numerically. However, we can set a precise bound by
requiring that the saddle point never be at $\gamma_U=0$
(where~(\ref{eq:orbavnoise}) diverges for all $D>0$). The applied
current at which such a saddle point occurs has already been derived
in~(\ref{eq:IC0}), leading us to the lower bound inequality
\be
\label{eq:lbound}
I\gg I_C^0.
\ee
as the validity condition for CEOA in thermally activated
scenarios. The reason for limiting ourselves to applied
currents greater than $I_C^0$ in deriving the upper bound
condition~(\ref{eq:IM}) is now justified.

As a result of~(\ref{eq:lbound}), there are only two scenarios in
which thermally activated switching can be analyzed and understood
using CEOA dynamics.  The first is in the limit $D\to0$ (the uniaxial
macrospin limit) where one has $\gamma(\epsilon)\to 1$ indpendently of
$\epsilon$. Divergences in the intensity of random contributions are
then avoided for all $\epsilon$ and~(\ref{eq:orbavnoise}) takes the
simple form $4\pi\sqrt{(\alpha/\xi)(1-\epsilon)}$. At $\epsilon=0$ it
converges to the finite value $4\pi\sqrt{\alpha/\xi}$ (which may still
be large, however, depending on the ratio $\alpha/\xi$).  The second
scenario is for models with $D<D_0$ and applied current values
$I>I_C^0$. In these, the saddle point to be reached via thermal
activation shifts to nonzero values $\gamma_U\ne 0$ where
(\ref{eq:orbavnoise}) does not diverge and the validity condition may
be satisfied.

The above discussion demonstrates that thermally activated switching
between dynamical limit cycle equilibria cannot be modeled with our
technique. Thermal switching from a limit cycle, in fact, must proceed
via noise diffusion up to a saddle point of the type $\gamma_U=0$
where noise contributions diverge. Nonetheless, it should be noted
that this does not necessarily invalidate the conditions for their
existence discussed in the previous section. In fact, relaxation to a
limit cycle is a drift-driven process for which noise related
contributions average to zero. Our analysis merely shows that, once
the macrospin has relaxed to its limit cycle state, further dynamics
must proceed via thermal activation for which CEOA does not provide a
suitable description.

In Fig.~7, the dynamical behavior for $D=4$ is displayed.
Here $D<D_0$ and $I_C^0<I_C^1$.  Fig.~7a shows the dependence between
rescaled current and mean switching time for different polar angular
tilts $\theta$ and fixed azimuthal angle $\phi$. For reference, the
critical currents $I_C^0$, $I_C^1$ and $I_C^M$ are shown. For large
applied currents, deviations between the different sets of data are
clearly visible. This discrepancy decreases as the current is lowered.

However, for $I<I_C^0$, the procedure is expected to fail due to the
previously discussed divergence of the orbital period. To show this,
we calculate the maximum deviation of the mean switching time between
all the angular data points at each value of the rescaled current (see
Fig.~7b).  As a further test, we have also numerically computed the
LHS of~(\ref{eq:validity}) and compared it to the angular
deviations. The critical currents $I_C^0$, $I_C^1$ and $I_C^M$ are
again shown for reference. The theory is in satisfactory agreement
with numerical simulations as can be seen from both the good alignment
of the deviation minimum with the minimum of the maximum flow and the
rapid spike in deviation as $I_C^0$ is approached.

\begin{figure}
\centering
\mbox{\subfigure{\includegraphics[clip,width=.5\textwidth, angle=0]{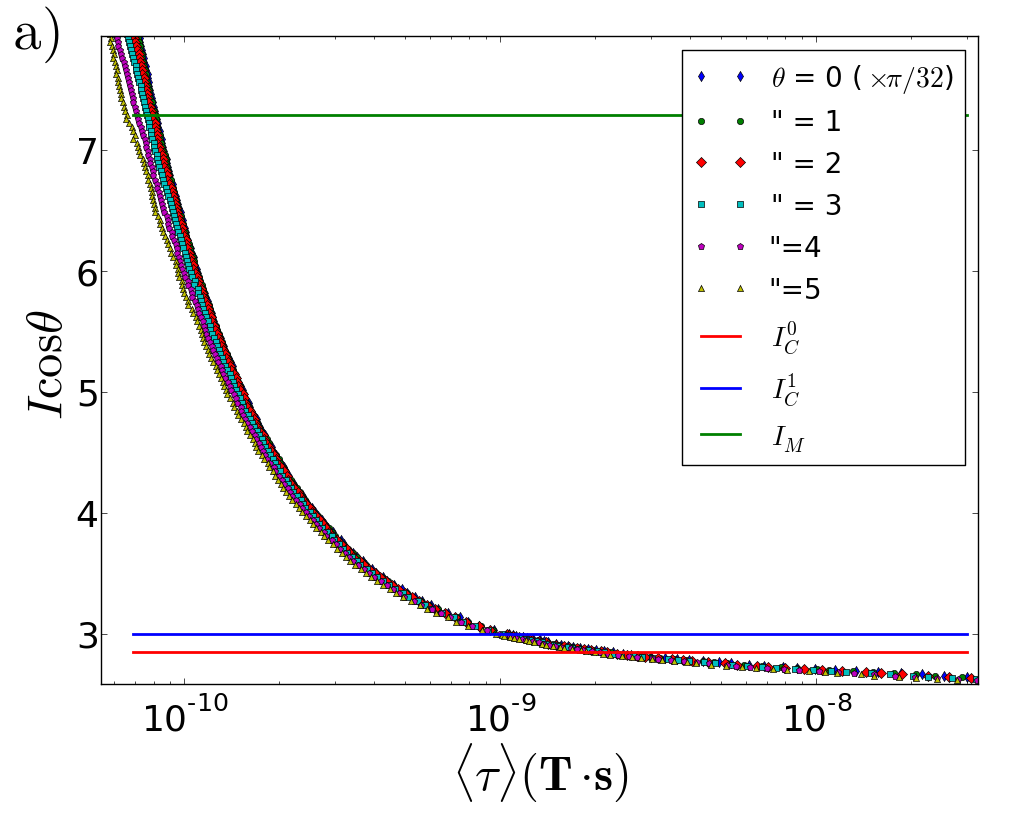}}\quad
\subfigure{\includegraphics[clip,width=.5\textwidth, angle=0]{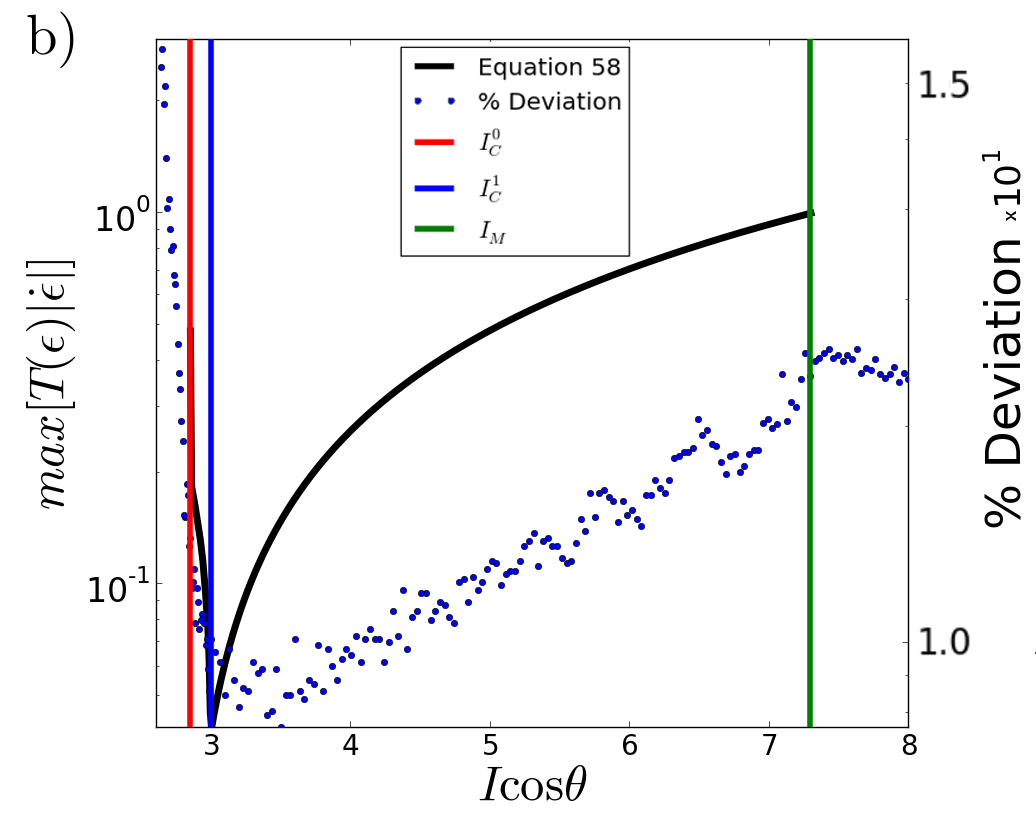}}}
\caption{{\bf a)} Mean switching time versus current for $D=4$, $\alpha=0.04$ and $\xi=80$ at different $\theta$ angular tilts with $\phi=0$ kept fixed. All currents have been rescaled by $1/\cos{\theta}$. Times are shown in units of ($s\cdot T$) where $T$ stands for Tesla: real time is obtained upon division by $H_K$. For visual guidance, the critical currents $I_C^0$ and $I_C^1$ have been included. In a regime where the CEOA technique is applicable, the switching data from the various angular configurations should all fall on top of each other. {\bf b)} Double axis plot of $\mathrm{max}[T(\epsilon)|\partial_{\mathrm{t}}|\epsilon||]$ and the percent deviation of data from ({\bf a)}) as a function of normalized current. In the current range where the deterministic flow achieves its minimum, the deviation of the data does also. As the critical current $I_C^0$ is approached, deviation spikes are observed analogously to what can be inferred by the theory.}
\label{F7}
\end{figure}

This discussion implies that the CEOA tecnique applies best to
small~$D$ macrospin models, and becomes increasingly inaccurate for
larger~$D$. Figs.~8 and~9 show the same analysis of a macrospin with
$D=50$ and one with $D=7$. Fig.~8a indicates that the technique is not
applicable for sufficiently high values of $D$. Nonetheless, theory and experiment approach each
other near the minimum of the deterministic flow, as seen in
Fig.~8b. As in Fig.~7, as currents drop below $I_C^0$, deviations in
the data rapidly spike.

The analysis and its comparison to numerical data indicate that the
CEOA tecnique is best suited for studying macrospin dynamics in the
crossover regime separating ballistic from fully activated thermal
switching. Furthermore, the stability analysis of the energy
dynamics~(\ref{eq:energydiff}) manages to capture the emergence of
limit cycles even though it fails to describe thermally activated
processes proceeding from them.

\begin{figure}
\centering
\mbox{\subfigure{\includegraphics[clip,width=.5\textwidth, angle=0]{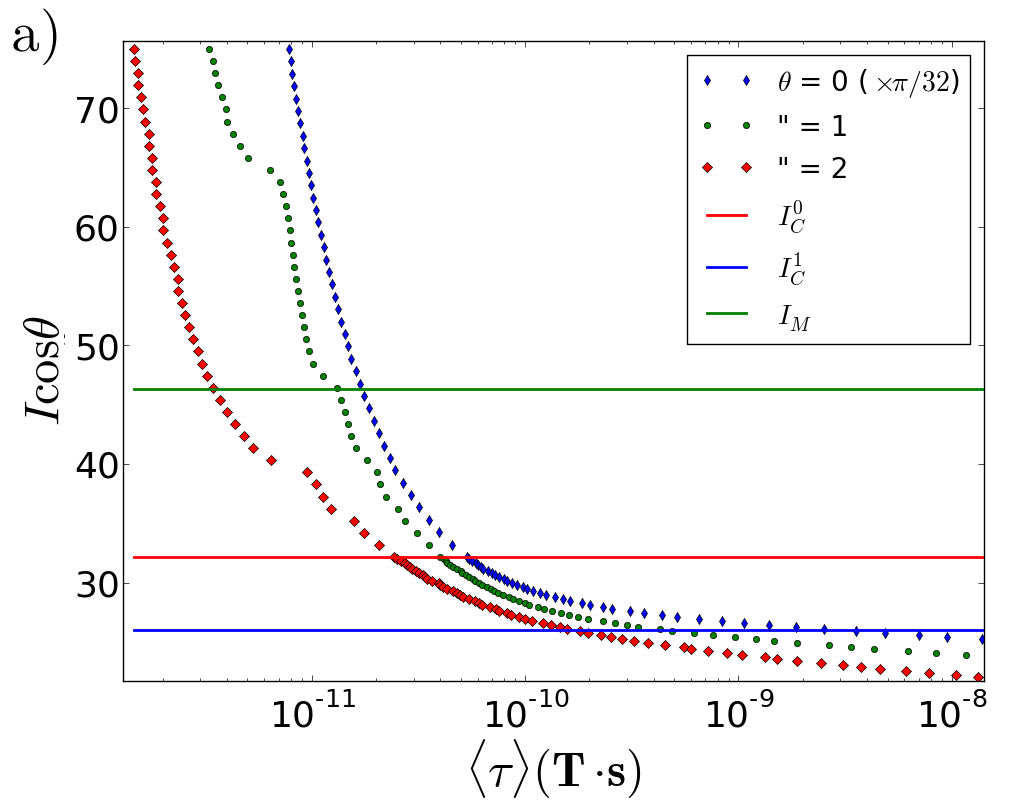}}\quad
\subfigure{\includegraphics[clip,width=.5\textwidth, angle=0]{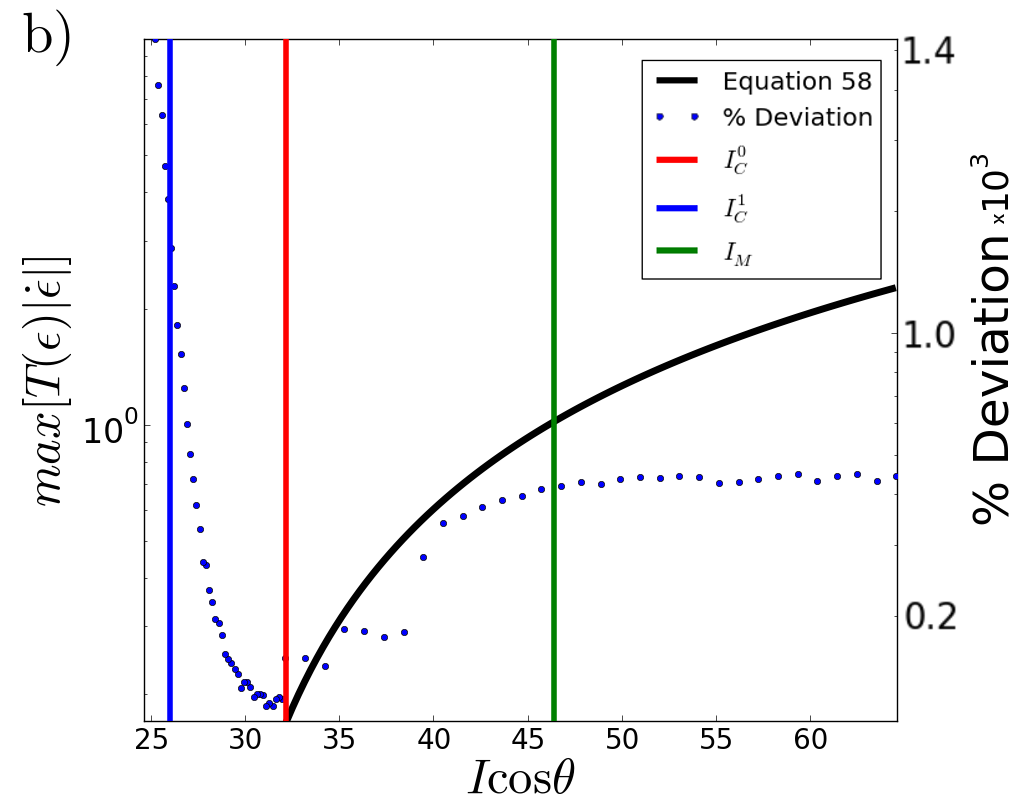}}}
\caption{{\bf a)} Mean switching time versus current for $D=50$, $\alpha=0.04$ and $\xi=80$ at different $\theta$ angular tilts with $\phi=0$ kept fixed. All currents have been rescaled by $1/\cos{\theta}$. Times are shown in units of ($s\cdot T$) where $T$ stands for Tesla: real time is obtained upon division by $H_K$. For visual guidance, the critical currents $I_C^0$ and $I_C^1$ have been included. {\bf b)} Double axis plot of $\mathrm{max}[T(\epsilon)|\partial_{\mathrm{t}}|\epsilon||]$ and the percent deviation of data from ({\bf a)}) as a function of normalized current. In the current range where the deterministic flow achieves its minimum, the deviation of the data does also. As the critical current $I_C^0$ is approached, deviation spikes are observed analogously to what can be inferred by the theory.}
\label{F8}
\end{figure}

\begin{figure}
\centering
\mbox{\subfigure{\includegraphics[clip,width=.5\textwidth, angle=0]{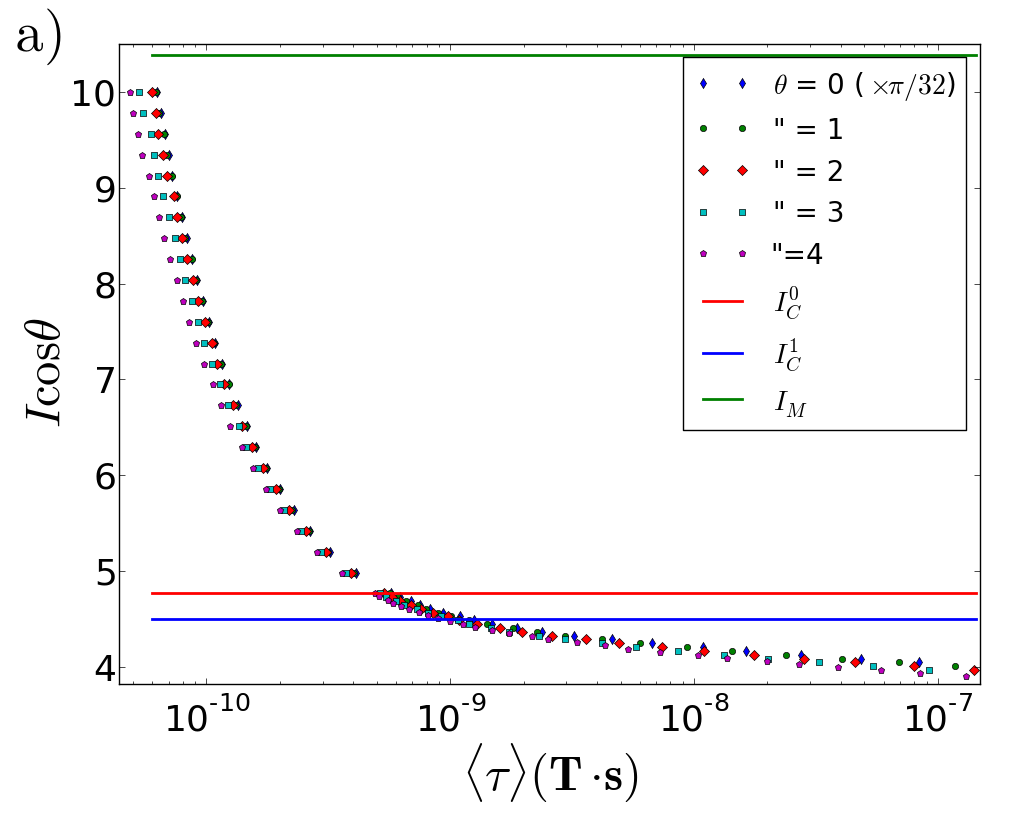}}\quad
\subfigure{\includegraphics[clip,width=.5\textwidth, angle=0]{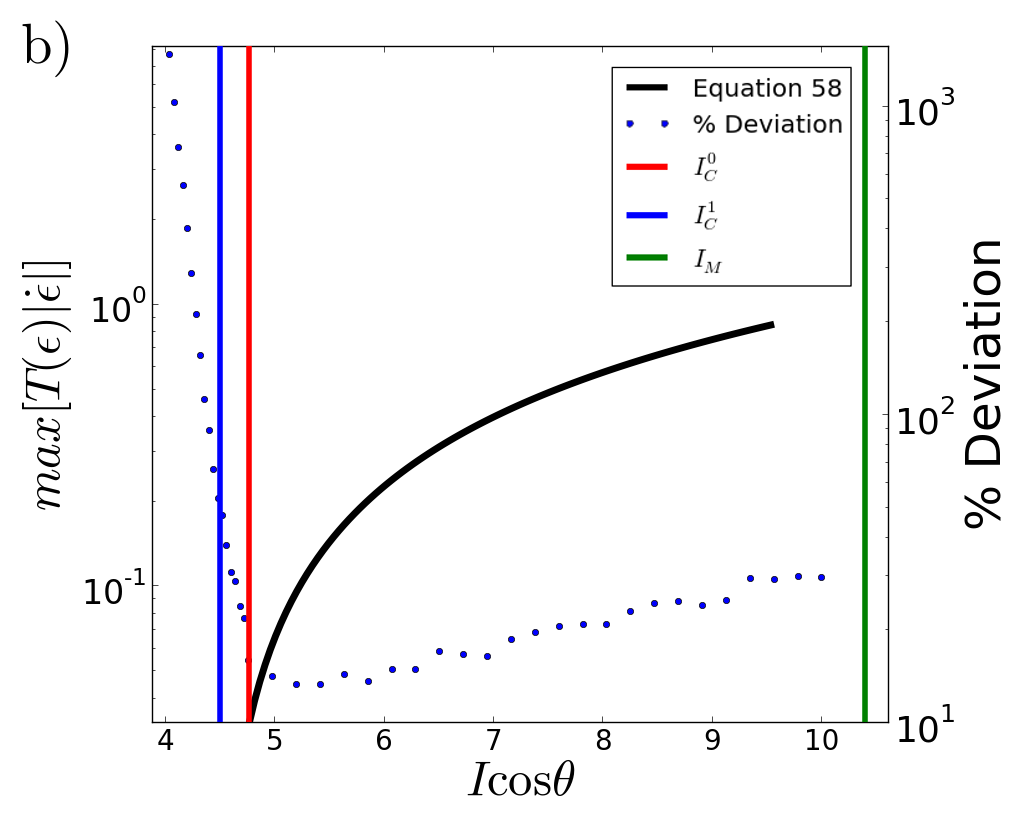}}}
\caption{{\bf a)} Mean switching time versus current for $D=7$, $\alpha=0.04$ and $\xi=80$ at different $\theta$ angular tilts with $\phi=0$ kept fixed. All currents have been rescaled by $n_z=1/\cos{\theta}$. Times are shown in units of ($s\cdot T$) where $T$ stands for Tesla: real time is obtained upon division by $H_K$. For visual guidance, the critical currents $I_C^0$ and $I_C^1$ have been included. {\bf b)} Double axis plot of $\mathrm{max}[T(\epsilon)|\partial_{\mathrm{t}}|\epsilon||]$ and the percent deviation of data from ({\bf a)}) as a function of normalized current. In the current range where the deterministic flow achieves its minimum, the deviation of the data does also. As the critical current $I_C^0$ is approached, deviation spikes are observed analogously to what can be inferred by the theory.}
\label{F9}
\end{figure}

\section{Thermally Activated Switching}

Scenarios in which thermally activated switching can be studied using
the approach described above are generally rather
limited. Nonetheless, in the previous section, the technique was shown
to be applicable in models with $D<D_0$ for currents
$I_C^0<I<I_C^1$. Starting from~(\ref{eq:energydiff}) one could in
principle construct a Fokker-Planck equation describing energy
diffusion and then attempt to solve an appropriate mean first passage
time problem numerically. Instead, we will simplify the thermal
activation problem by deriving the exponential scaling dependence
between mean switching time and current, using an analytical tool
described in~\cite{MMM}. In that paper, it was shown how, starting
from energy diffusion dynamics analogous to~(\ref{eq:energydiff}), the
exponential scaling dependence can be written in terms of an integral
from the initial stable state to the saddle using a
Friedlin-Wentzell~\cite{FW} type formulation: 
\be
\log(\langle\tau\rangle)\propto2\int_{|\epsilon_U|}^1\frac{f(\epsilon')}{g^2(\epsilon')}d\epsilon'
\ee 
where $f(\epsilon)$ and $g(\epsilon)$ are, respectively, the drift and
noise terms of~(\ref{eq:energydiff}), now expressed in terms of the
energy~$\epsilon$. This integral can be computed numerically once the
saddle point of the energy flow has been identified (e.g. via a Newton
algorithm). The integrand can be further expanded for $D>>1$ and
approximated around the stable point $|\epsilon|=1$. This gives, to
first order in $1-|\epsilon|$,
\bea
\label{eq:expscaling}
\log(\langle\tau\rangle)&\propto&\xi\int_{|\epsilon_U|}^{1}\left(1-(I/I_C^0)\sqrt{\frac{D}{D+\epsilon'}}\left(\frac{1-\epsilon'}{\eta_1(\epsilon')-\epsilon'\eta_0(\epsilon')}\right)\right)d\epsilon'\\
&\simeq&\xi\int_{|\epsilon_U|}^{1}\left(1-(I/I_C^0)\frac{1-\epsilon'}{1-\epsilon'^{\zeta(D)}}\right)d\epsilon'
\eea
where $\zeta(D)$ is given by
\be
\zeta(D)=\frac{I_C^1}{I_C^0}=\frac{\pi}{4}\left(\frac{D+2}{\sqrt{D(D+1)}}\right).
\ee
The approximated integral can now be written in closed form in terms
of hypergeometric functions. The resulting expression is accurate to
within $2\%$ for values $D>0.1$.

In Fig.~10 we plot the dependence of the mean switching time as a
function of applied current, and compare it to
$\log(\tau)\propto\xi(1-I/I_C)^{\beta}$. In the limit $D\to0$, the
uniaxial switching exponent $\beta=2$ is recovered. For larger values
of $D$, the exponent $\beta$ depends non-linearly on the applied
current $I$. In the limit $I\to 0$ the limiting value of $\beta$ can
be derived analytically:
\be
\label{eq:betalim}
\lim\limits_{I\to0}\beta(I)=\lim\limits_{I\to0}\frac{\log(1-\frac{q(D)}{I_C^0}I)}{\log(1-I/I_C^1)}=q(D)\frac{I_C^1}{I_C^0},
\ee
with
\be
q(D)\equiv\int_0^1\sqrt{\frac{D}{D+x}}\left(\frac{1-x}{\eta_1(x)-x\eta_0(x)}\right)dx.
\ee

The same calculation can be performed in the limit $I\to I_C^1$ and
used to show that the exponent $\beta$ diverges for all non-zero
values of $D$. The mean switching time is nonetheless well behaved, as
can be seen in Fig.~10a. This differs from the results obtained by
Taniguchi~\cite{TaniguchiE}, in which the limiting value of the
exponent $\beta$ is roughly $2.2$, as the applied current approaches
the critical threshold for determinisic switching.

\begin{figure}
\centering
\mbox{\subfigure{\includegraphics[clip,width=.5\textwidth, angle=0]{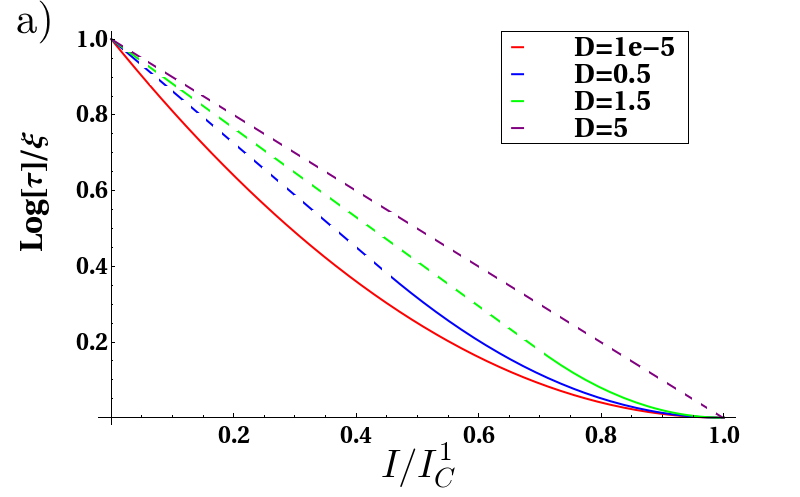}}\quad
\subfigure{\includegraphics[clip,width=.5\textwidth, angle=0]{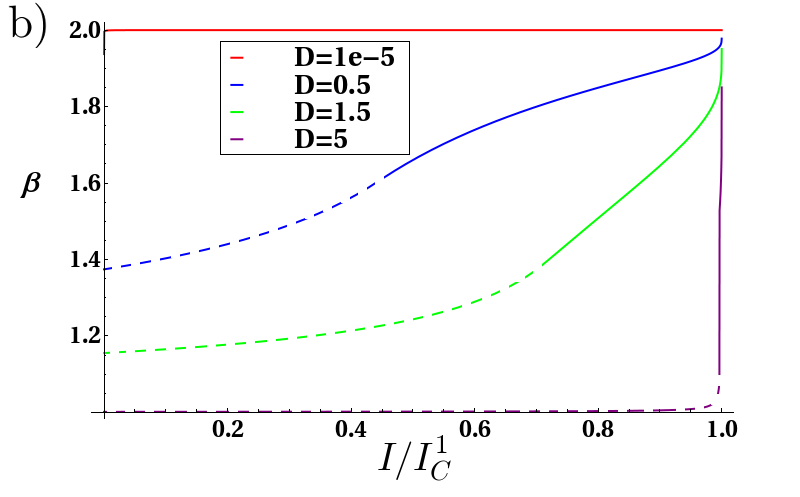}}}
\caption{{\bf a)} Scaling dependence of mean switching time as a function of applied current $I$ for models with varying $D<D_0$. $\xi$ is the energy barrier height and $I_C^1$ the critical current threshold for deterministic switching. {\bf b)} Fit of (\ref{eq:expscaling}) to the form $(1-I/I_C^1)^{\beta}$. Dashed lines represent continuation of analytical results outside the tecnique's regime of validity. Fitting exponent $\beta$ is plotted as a function of applied current for models with varying $D$. In the limit of small $D$ the exponent approaches the constant value $\beta=2$ consistent with previous uniaxial macrospin results~\cite{Taniguchi,APL,MMM}. For $D>0$, the exponent $\beta$ depends nonlinearly on the applied current intensity. Only for values $D\sim D_0$ do we notice that in the limit of small applied currents, $\beta\to 1$ as suggested by similar energy diffusion studies from the literature~\cite{Apalkov,Katie}. For intermediate values $D_0>D>0$ the low current limit of $\beta$ can be obtained analytically (\ref{eq:betalim}). On the other hand, in the limit $I\to I_C^1$ the exponent $\beta$ can be shown to diverge for all non-zero values of $D$.}
\label{F10}
\end{figure}

\section{Conclusion}

This paper extends the analytical approach introduced by the authors
in~\cite{MMM} and applies it to the biaxial macrospin model. We have
described a technique that reduces the complexity of the
$3D$~macrospin reversal dynamics, under the action of both spin-torque
and thermal noise, to a $1D$~stochastic equation in the energy space of
the macrospin. To do so, the complete macrospin dynamics were
approximated over precessional constant energy orbits.

Under such an approximation, the resulting theory predicts that the
geometries involved influence the respective dynamics in very precise
ways. We found that the angular tilt between spin polarization and
uniaxial anisotropy axes factors into the dynamics only as a trivial
rescaling of the applied current. Similarly, the relative orientation
of the hard axis is predicted to play no role in the switching
dynamics under conditions in which constant-energy orbit averaging
(CEOA) is a valid assumption.

The main parameter characterizing the different dynamical regimes is
the ratio $D$ between the hard and easy axis anisotropies. For a
specific value of~$D$, two critical currents ($I_C^1=(D+2)/2$,
$I_C^0=(2/\pi)\sqrt{D(D+1)}$) are found to exist. The relative
magnitude of these critical currents depends nonlinearly on~$D$. For
$D>D_0\simeq5.09$, we showed that stable limit cycle magnetization
precessions appear in well-defined current ranges; transitions between
these stable limit cycles proceeds through thermal activation.
When~$D<D_0$, limit cycles generally do not appear and the switching
dynamics become qualitatively similar to those of a uniaxial macrospin
model.

The CEOA procedure requires that the energy diffusion due to spin
torque and thermal noise take place on a timescale much larger than
the conservative precessional timescale of the model. This allows us to
establish general conditions for the validity of CEOA dynamics.

We test our findings numerically by solving the full macrospin
evolution equations and analyzing their applied current vs.~mean
switching time curves for various angular configurations. Our
techniques are found to be applicable in explaining ranges of applied
currents such that: $I\gg I_C^0$ where any thermally activated process
proceeds along nearly precessional trajectories, and $I/I_C^0\ll
1+1/(8\alpha\sqrt{D})$ such that spin torque intensity is minimized
along the deterministic flow. These conditions do not generally hold
in large~$D$ models, but do for small~$D$. In the limit~$D\to 0$ we
recover known results for the uniaxial macrospin model. Our technique
also appears to be suitable for studying thermally activated switching
in models with~$D<D_0$.

Finally, Friedlin-Wentzell theory was employed to analytically study
the exponential scaling dependence between mean switching time and
applied current in thermally activated scenarios. The exact analytical
scaling was reduced to quadratures and an analytical approximation was
suggested.  The resulting analytical scaling dependence to the
standard form $\log(\tau)\propto\xi(1-I/I_C)^{\beta}$ and the current
dependence of the exponent $\beta$ was studied.

The exponent~$\beta$ was found to depend nonlinearly on the applied
current intensity, similar to~\cite{TaniguchiE}.  In the uniaxial
macrsopin limit~$D\to0$, the constant $\beta=2$ result was recovered.
For~$D\ne0$, $\beta$ was found to depend on both the applied current
intensity and the precise value of~$D$.

\begin{acknowledgments}

  The authors would like to acknowledge A. MacFadyen and J. Z. Sun for
  useful discussions and comments leading to this paper. This research
  was supported by NSF-DMR-100657, NSF-DMR-1309202 and PHY0965015.

\end{acknowledgments}

\appendix
\section{Elliptic Integral Identities}

The identities

\bea
\mathrm{e}(x)\equiv x\mathrm{E}(1-\frac{1}{x^2})=\mathrm{E}(x)\\
\mathrm{k}(x)\equiv \frac{1}{x}\mathrm{K}(1-\frac{1}{x^2})=\mathrm{K}(x)
\eea
can be immediately proven by considering the defining differential equation satisfied by $K(x)$, namely~\cite{Whittaker}:

\be
\partial_x\mathrm{K}(x)=\frac{\mathrm{E}(x)-(1-x)\mathrm{K}(x)}{2x(1-x)}.
\ee

Computing then explicitely the derivative of $k(x)$ using definition (A1) and rearranging, one finds that:

\be
\partial_x\mathrm{k}(x)=\frac{\mathrm{e}(x)-(1-x)\mathrm{k}(x)}{2x(1-x)},
\ee
in other words they satisfy the same differential relation.

\end{document}